# Adaptive Backstepping Chaos Synchronization of Fractional order Coullet Systems with Mismatched Parameters

M. Shahiri T.* , A. Ranjbar **
R. Ghaderi*, M. Karami*, S.H. Hoseinnia***

*Noshirvani University of Technology, Faculty of Electrical and Computer Engineering, Intelligent system Research Group, P.O. Box 47135-484, Babol, IRAN (e-mail: ma_shairi@stu.nit.ac.ir, r_ghaderi@yahoo.co.uk , mkarami@nit.ac.ir ).
** Golestan University, P.O. Box 155, Post Code: 49138-15759, Gorgan, Iran (e-mail: a.ranjbar@gu.ac.ir).
*** Dept. of Electrical, Electronic and Automation Engineering, Industrial Engineering School, University of Extremadura, Badajoz, Spain, (e-mail: shossein@alumnos.unex.es)}.

**Abstract:** In this paper, synchronization of fractional order Coullet system with precise and also unknown parameters are studied. The proposed method which is based on the adaptive backstepping, has been developed to synchronize two chaotic systems with the same or partially different attractor. Sufficient conditions for the synchronization are analytically obtained. There after an adaptive control law is derived to make the states of two slightly mismatched chaotic Coullet systems synchronized. The stability analysis is then proved using the Lyapunov stability theorem. It is the privilege of the approach that only needs a single controller signal to realize the synchronization task. A numerical simulation verifies the significance of the proposed controller especially for the chaotic synchronization task.

*Keywords*: Fractional order Differential Equations (FDEs), Chaos, Coullet system, Synchronization, Adaptive Backstepping, Nonlinear control.

## 1. INTRODUCTION

Fractional calculus is an old mathematical topic, which has been originated from 17th century. Nowadays, some fractional order differential systems such as Chua circuit (Hartley et al. 1995), Duffing system (Caponetto et al.1997) jerk model (Ahmad et al., 2003), Chen system (Lu et al., 2006a [4]), the fractional order Lü (Lu ,2006b), Rossler (Li et al.,2004), Arneodo (Lu ,2005) and Newton–Leipnik systems (Sheu et al., 2008) have been found useful to demonstrate chaotic behaviours. Sensitivity to initial conditions is an important characteristic of chaotic systems. Therefore, chaotic systems are usually difficult to be controlled or synchronized. Control of these systems has been considered as an important and challenging problem (Liu et al., 2008).
Control of chaotic systems would have been supposed impossible with uncontrollable and unpredictable dynamic.
The problem of designing a system, whose behaviour mimics that of another chaotic system, is called synchronization. The combination of two chaotic systems is usually called drive (master) and response (slave) systems. Different control technique *e.g.* a chattering-free fuzzy sliding-mode control (FSMC) strategy for synchronization of chaotic systems even in presence of uncertainty has been proposed in (Deng et al., 2005a). In (Deng et al., 2005b, Delavari et al., 2010) authors have proposed an active sliding mode control to synchronize two chaotic systems with parametric uncertainty. Over the past decade, Backstepping has become the most popular design procedure for adaptive nonlinear control. It is because it guarantees the global stabilities, tracking, and transient performance for a broad class of strict feedback systems. In (Zhang , 2004, Yu, 2004, Tan, 2003), it has been shown that many well-known chaotic systems as paradigms in research of chaos, including Duffing oscillator, Van der Pol oscillator, Rossler, Lorenz, Lü, Chen systems and several type of Chua's circuit, can be transformed into a class of nonlinear system in the so-called non-autonomous form. Meanwhile backstepping and tuning functions control schemes have been employed and also extended to control these chaotic systems with key parameters unknown. In particular, the controlled chaotic system has been demanded to asymptotically track a smooth and bounded reference signal which generated from a known reference (chaotic) model. Recently some work has been reported for synchronization of fractional order Coullet system via nonlinear design method (Sahiri et al., 2008, 2009, 2010), but there is lack of report of backstepping control of fractional order Coullet system.
In this paper, a backstepping approach is designed for synchronization task. Meanwhile a novel adaptive backstepping scheme has been proposed based on the Lyapunov stability theory for chaotic synchronization of fractional order Coullet system with three unknown parameters.
The paper is organized as follows:
Section 2 summarizes chaos in fractional order Coullet systems, regarding to some results. In Section 3 Backstepping technique is extended to synchronize a fractional order Coullet (with both nominal and unknown parameters). Simulation approach is also given in two different cases in Section 4. Ultimately, the work will be concluded in section 5.

## 2. CHAOS IN FRACTIONAL ORDER COULLET SYSTEM

In advance a brief description of the stability of fractional order systems will be expressed.

*2.1. Stability analysis of Fractional order systems*

A fractional order linear time invariant model can be represented by the following state-space format:

$$\begin{cases} D^q x = Ax + Bu \\ y = Cx \end{cases} \quad (1)$$

where, $x \in \mathbb{R}^n$, $u \in \mathbb{R}^r$ and $y \in \mathbb{R}^p$ denote states, input and output vectors of the system, respectively. Corresponding coefficients will be shown by matrices $A \in \mathbb{R}^{n \times n}$, $B \in \mathbb{R}^{n \times r}$ and $C \in \mathbb{R}^{p \times n}$ respectively, whilst $q$ is a real fractional commensurate order. Fractional order differential equations are at least as stable as their integer orders counterparts. This is because; systems with memory are typically more stable than their memory-less alternatives (Ahmed et al, 2007). It has been shown that an autonomous dynamic $D^q x = Ax$, $x(0) = x_0$ is asymptotically stable if the following condition is met (Matignon, 1996):

$$|\arg(eig(A))| > q\pi/2, \quad (2)$$

where $0 < q < 1$ and $eig(A)$ represents eigenvalues of matrix $A$. Furthermore, the system is stable if $|\arg(eig(A))| \geq q\pi/2$ whilst those critical eigenvalues which satisfy $|\arg(eig(A))| = q\pi/2$ have geometric multiplicity of one. The stability region for $0 < q < 1$ are shown in Fig. 1. Consider the following autonomous commensurate fractional order system:

$$D^q x = f(x), \quad (3)$$

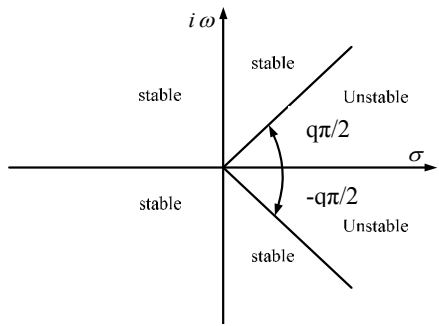

Fig. 1: Stability region of the LTI FO system with $0 < q < 1$

where, $0 < q < 1$ and $x \in \mathbb{R}^n$. Equilibrium points of system (3) are calculated when:

$$f(x) = 0. \quad (4)$$

These points are locally asymptotically stable if all eigenvalues of Jacobian matrix $J = \partial f / \partial x$, which are evaluated at the equilibrium points- satisfy the following condition (Ahmed et al,2007; Matignon,1996):

$$|\arg(eig(J))| > q\pi/2, \quad (5)$$

*2.2. Chaos in Fractional Order Coullet system*

Some of nonlinear systems represent a chaotic behaviour. These systems are sensitive to initial conditions. This means two identical systems but with a minor deviation in their initial condition may approach to a completely different result. This means having known bounded initial conditions, there is less chance to predict the dynamic behaviour. This deterministic treatment is called chaos. In this part, chaotic Coullet system will be studied. The system is consisted fractional derivatives, which are as follows:

$$\begin{cases} D^q y_1 = y_2 \\ D^q y_2 = y_3 \\ D^q y_3 = cy_3 + by_2 + ay_1 + dy_1^3 + u(t) \end{cases} \quad (6)$$

The response is evaluated when parameters of the system are choosing as (Hu *et. al.*, 2008): $a = 0.8, b = -1.1, c = -0.45$ and $d = -1$. Eigenvalues of Jacobin matrix of Coullet system at the equilibrium points are spotted as follows:

$$\begin{cases} (0.5054, -0.4777 \pm 1.1639i) \\ (-0.9842, 0.2671 \pm 1.2468i) \end{cases}$$

Eigenvalues show that first equilibrium point does not meet the necessary condition to exhibit chaos and only the second one can be responsible for the generation of scrolls (Shahiri et al., 2010).

These yield the stability region according to the following boundary of the fractional parameters $q$ using equation (5) as:

$$\begin{cases} |\arg(0.2671 \pm 1.2468i)| = \\ 1.3598 > q\pi/2 \rightarrow q < 0.8656 \end{cases} \Rightarrow q < 0.8656$$

This means for $q > 0.8656$ the system is chaotic, whereas the other bound provides stable region. In this system, the chaotic threshold will be as $3 \times 0.8656 = 2.60$. Fig.2 shows a simulation results concerning to different value of $q$ (Shahiri et al., 2010). Graphs for $q = 0.98, 0.95, 0.87$ show chaotic behaviour for the system which satisfy the theoretical inspection. This also confirms the outcome for $q = 0.86$ which explicitly shows the stability of the system.

## 3. BACKSTEPPING CONTROL OF FRACTIONAL ORDER CHAOTIC SYSTEM

Backstepping design has been recognized as a powerful method to control and synchronize chaos (Yu, 2004, Tan, 2003, Yassen, 2006, 2007, Wang, 2007, Peng, 2008). It has been reported (Yu, 2004, Tan, 2003, Yassen, 2006, 2007, Wang, 2007, Peng, 2008) that backstepping design can guarantee global stability whilst promotes transient performance for a broad class of strict-feedback nonlinear systems. Eventually, backstepping method is a systematic Lyapunov based control technique, which can be applied to strict, pure-feedback and block-strict-feedback systems (Mascolo et al., 1999). However a novel adaptive backstepping scheme will be proposed based on the Lyapunov stability theory for chaotic synchronization of fractional order Coullet system with three mismatch parameters.

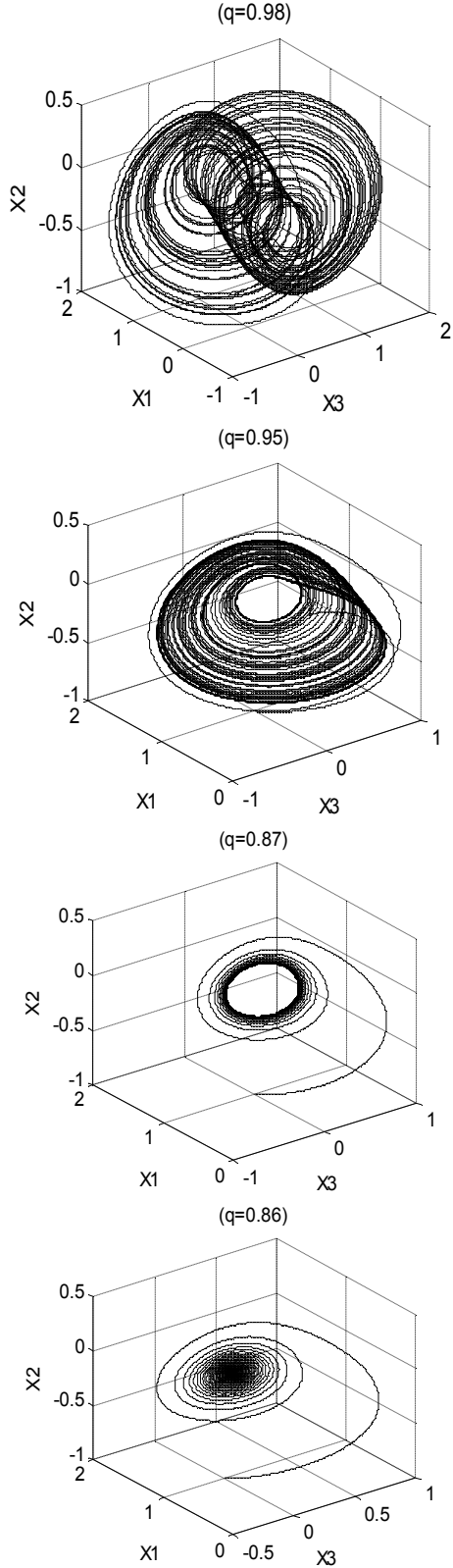

Figure 2: Phase portrait of chaotic Coullet system for different values of the fractional parameter

### 3.1. Synchronization of two Coullet systems with nominal parameters

In this section, the control procedure will be used to synchronize two identical fractional order Coullet systems with different initial conditions. An open loop response of two identical chaotic systems with different initial condition $(x_{10}, x_{20}, x_{30}) \neq (y_{10}, y_{20}, y_{30})$ will quickly deviate. However, using appropriate control scheme may achieve synchronization for any bounded initial value. Consider the following fractional order Coullet system as a master:

$$\begin{cases} D^q x_1 = x_2 \\ D^q x_2 = x_3 \\ D^q x_3 = cx_3 + bx_2 + ax_1 + dx_1^3 \end{cases} \quad (7)$$

The objective of this paper is to design a controller $u(t)$ such that the controlled slave system of the form:

$$\begin{cases} D^q y_1 = y_2 \\ D^q y_2 = y_3 \\ D^q y_3 = cy_3 + by_2 + ay_1 + dy_1^3 + u(t) \end{cases} \quad (8)$$

becomes synchronous with the master chaotic system (7). The error is defined as the discrepancy of the relevant states i.e. $e_i = y_i - x_i$ for $i = 1, 2, 3$. Deducing the master dynamic from that of the slave leads to:

$$\begin{cases} D^q e_1 = e_2 \\ D^q e_2 = e_3 \\ D^q e_3 = ce_3 + be_2 + ae_1 + d(y_1^3 - x_1^3) + u(t) = \\ ce_3 + be_2 + ae_1 + de_1 y_1^2 + de_1^2 x_1 + 2de_1 x_1^2 + u(t) \end{cases} \quad (9)$$

In continuing the backstepping procedure will be introduced to design a synchronizing control signal $u(t)$.

**Theorem 1.** The controller $u(t)$ in (10) achieves the synchronization:

$$u(t) = -\left(3(D^{q-1})^3 + a + dy_1^2 + dx_1 e_1 + 2dx_1^2\right) e_1 \\ -\left(5\left(D^{q-1}\right)^2 + b\right) e_2 - \left(3(D^{q-1}) + c\right) e_3 \quad (10)$$

**Proof**: Primarily, the stability of the first equation in (9) will be shown. Let us define a dummy variable $w_1 = e_1$, then the derivative will be obtained as:

$$D^q w_1 = D^q e_1 = e_2 \quad (11)$$

A Lyapunov function is candidate as:

$$V_1(w_1) = \frac{1}{2} w_1^2 > 0 \quad (12)$$

The time derivative is immediately obtained as:

$$\dot{V_1}(w_1) = e_1 \cdot \dot{e_1} = e_1 D^{1-q}(e_2) = -w_1^2 + w_1(e_1 + D^{1-q}(e_2)) \quad (13)$$

Secondly, select $w_2 = e_1 + D^{1-q}(e_2)$ and candidate the second Lyapunov function as:

$$V_2(w_1, w_2) = V_1(w_1) + \frac{1}{2}w_2^2 > 0 \quad (14)$$

By some manipulation, one can verify that the time derivative of $V_2$ is given by:

$$\dot{V_2}(w_1, w_2) = -w_1^2 - w_2^2 + w_2(2e_1 + 2D^{1-q}e_2 + (D^{1-q})^2 e_3) \quad (15)$$

Ultimately, the last Lyapunov function may be formed as:

$$V_3(w_1, w_2, w_3) = V_2(w_1, w_2) + \frac{1}{2}w_3^2 > 0 \quad (16)$$

where $w_3 = 2e_1 + 2D^{1-q}e_2 + (D^{1-q})^2 e_3$. The time derivative of $V_3$ is:

$$\dot{V_3}(w_1, w_2, w_3) = -w_1^2 - w_2^2 - w_3^2 + w_3\left(3e_1 + 5D^{1-q}e_2 + 3(D^{1-q})^2 e_3 + (D^{1-q})^3 (A + u(t))\right) \quad (17)$$

where:

$$A = ce_3 + be_2 + ae_1 + de_1 y_1^2 + de_1^2 x_1 + 2de_1 x_1^2 \quad (18)$$

Therefore, the control input (10) provides:

$$\dot{V_3}(w_1, w_2, w_3) = -w_1^2 - w_2^2 - w_3^2 = -2V_3 < 0 \text{ in } D - \{0\}$$
and $\dot{V_3}(0) = 0$. $\quad (19)$

From (19) can be seen that $V_3 \leq V_3(0)e^{-2t}$ which shows an exponential stability i.e. $\lim_{t \to \infty} V_3(t) = 0$. This implies that: $\lim_{t \to \infty} w_i = 0$ for $i = 1, 2, 3$.:

$$\begin{cases} w_1 \to 0 \Rightarrow e_1 \to 0 \\ w_2 \to 0 \Rightarrow e_1 + D^{1-q}e_2 \to 0 \Rightarrow \dot{e_1} \to 0 \end{cases} \quad (20)$$

As $e_1, \dot{e_1} \to 0$, from the first and second formula in (9) one can result $e_2, e_3 \to 0$. Hence, the master synchronizes the slave by the controller in (10). This completes the proof. □

In the preceding an adaptive synchronization of Coullet system with unknown parameters will be considered.

### 3.2. Synchronization of two Coullet systems with unknown parameters

An adaptive synchronization of Coullet system with unknown parameters will be expressed here. Primarily consider the problem with only one unknown coefficient $c$. The problem will be extended to consider three unknown parameters thereafter. The aim is to design a signal $u(t)$ to control the salve Coullet system as:

$$\begin{cases} D^q y_1 = y_2 \\ D^q y_2 = y_3 \\ D^q y_3 = \hat{c}y_3 + by_2 + ay_1 + dy_1^3 + u(t) \end{cases} \quad (21)$$

where the parameter $\hat{c}$ denotes the estimation of parameter $c$. The designation of the controller $u(t)$ of:

$$u(t) = -\left(3(D^{q-1})^3 + a + dy_1^2 + dx_1 e_1 + 2dx_1^2\right)e_1 \\ -\left(5(D^{q-1})^2 + b\right)e_2 - \left(3(D^{q-1}) + \hat{c}\right)e_3 \quad (22)$$

together with an adaptation rule for $\hat{c}$ as:

$$\dot{\hat{c}} = -\left(2e_1 + 2D^{1-q}e_2 + (D^{1-q})^2 e_3\right)\left(D^{1-q}\right)^3 x_3 \quad (23)$$

proves the slave Coullet system to become synchronous with the master especially with an unknown parameter $c$.

**Proof.** The error between dynamics (7) and (21) is defined by:

$$\begin{cases} D^q e_1 = e_2 \\ D^q e_2 = e_3 \\ D^q e_3 = \hat{c}y_3 - cx_3 + be_2 + ae_1 + d(y_1^3 - x_1^3) + u(t) = \\ \hat{c}y_3 - cx_3 + be_2 + ae_1 + de_1 y_1^2 + de_1^2 x_1 + 2de_1 x_1^2 + u(t) \end{cases} \quad (24)$$

Candidate a Lyapunov function as:

$$V = \frac{1}{2}(w_1^2 + w_2^2 + w_3^2) + \frac{1}{2}(\hat{c} - c)^2 \quad (25)$$

where $w_1 = e_1$, $w_2 = e_1 + D^{1-q}(e_2)$ and $w_3 = 2e_1 + 2D^{1-q}e_2 + (D^{1-q})^2 e_3$. The time derivative of $V$ is as follows:

$$\dot{V} = -w_1^2 - w_2^2 - w_3^2 + \\ w_3\left[\left(3 + (D^{1-q})^3 (de_1 y_1^2 + de_1^2 x_1 + 2de_1 x_1^2 + a)\right)e_1 + \\ \left(5D^{1-q} + b(D^{1-q})^3\right)e_2 - 3(D^{1-q})^2 e_3 + (D^{1-q})^3 \hat{c}y_3 - \\ (D^{1-q})^3 cx_3 + (D^{1-q})^3 u(t)\right] + (\hat{c} - c)\dot{\hat{c}} \quad (26)$$

Substitution of $u(t)$ and the adaptation rule in $\dot{V}$ leads to:

$$\dot{V} = -w_1^2 - w_2^2 - w_3^2. \quad (27)$$

It also should remind that since a chaotic system has bounded trajectories, there exists a positive constant $L$ that $|x_i| < L, |y_i| < L (i = 1, 2, 3)$. Since $\dot{V} \leq 0$ one can result $e_1, D^{1-q}e_2, (D^{1-q})^2 e_3, \hat{c} \in L_\infty$. From (28)

$$\int_0^t \|w\|^2 dt = V(0) - V(t) \leq V(0) \quad (28)$$

This results that $w_1, w_2, w_3 \in L_2$ and $e_1, D^{1-q}e_2, (D^{1-q})^2 e_3 \in L_2$. By manipulating error dynamic in (24) immediately follows:

$$\begin{cases} \dot{e_1} = D^{1-q}e_2 \\ \dot{e_2} = D^{1-q}e_3 \\ \dot{e_3} = D^{1-q}\left(\hat{c}y_3 - cx_3 + be_2 + ae_1 + d(y_1^3 - x_1^3) + u(t)\right) \end{cases} \quad (29)$$

which show $\dot{e_1}, \dot{e_2}, \dot{e_3} \in L_\infty$. Therefore using the Barbalat's lemma implies that $e_1(t), e_2(t), e_3(t) \to 0$ as $t \to \infty$ (Shi et.al., 2009). Consequently it shows that the controller (22) and the update law (23) synchronize the slave with the master. This completes the proof. □

Similarly, the Coullet system will be extended to consider three mismatched coefficients. These unknown parameters *a, b,* and *c* and the controller will be updated as:

$$\dot{\hat{a}} = \left(2e_1 + 2D^{1-q}e_2 + \left(D^{1-q}\right)^2 e_3\right)\left(D^{1-q}\right)^3 x_1$$

$$\dot{\hat{b}} = -\left(2e_1 + 2D^{1-q}e_2 + \left(D^{1-q}\right)^2 e_3\right)\left(D^{1-q}\right)^3 x_2 \quad (30)$$

$$\dot{\hat{c}} = -\left(2e_1 + 2D^{1-q}e_2 + \left(D^{1-q}\right)^2 e_3\right)\left(D^{1-q}\right)^3 x_3$$

$$u(t) = -\left(3\left(D^{q-1}\right)^3 + \hat{a} + dy_1^2 + dx_1 e_1 + 2dx_1^2\right)e_1$$
$$-\left(5\left(D^{q-1}\right)^2 + \hat{b}\right)e_2 - \left(3\left(D^{q-1}\right) + \hat{c}\right)e_3 \quad (31)$$

where $V = \frac{1}{2}(w_1^2 + w_2^2 + w_3^2) + \frac{1}{2}(\hat{c}-c)^2 + \frac{1}{2}(\hat{b}-b)^2 + \frac{1}{2}(\hat{a}-a)^2$.

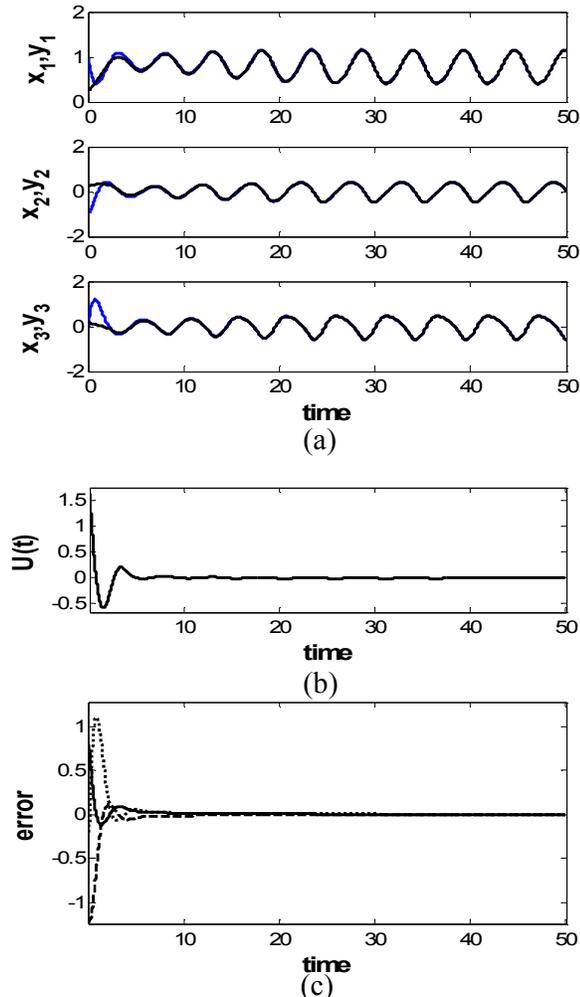

(a)

(b)

(c)

Fig. 3: a) Simulation Results of synchronization of two identical fractional order Coullet systems with nominal parameter $a = 0.8, b = -1.1, c = -0.45$ and $d = -1$ for $q = 0.9$ b) Control signal $u(t)$. c) The synchronization error.

## 4. NUMERICAL RESULT

In this section, to verify the effectiveness of the described method, a simulation approach for two identical fractional orders Coullet system are brought here.

**Case 1**. Initial conditions will be considered different as: $(x_{10}, x_{20}, x_{30}) = (1, -1, 0)$, and $(y_{10}, y_{20}, y_{30}) = (0.2, 0.2, 0.2)$. The parameters and the fractional order are respectively chosen as: $a = 0.8, b = -1.1, c = -0.45$ and $q = 0.9$. It is found that the Coullet system exhibits a chaotic behaviour. Synchronization of systems (7) and (8) via nonlinear control law (10) is illustrated in Figs. 3 which display the state responses, the control signal and synchronization error of systems, respectively.

**Case 2**. Consider the Coullet system (9) with two unknown parameters $b$, and $c$. Initial conditions of the master and the salve are the same as in case 1, when $q = 0.95$. From Fig. 4 can be seen that the adaptive control law (32) together with the adaptation law in (33) synchronizes the slave system to track the master.

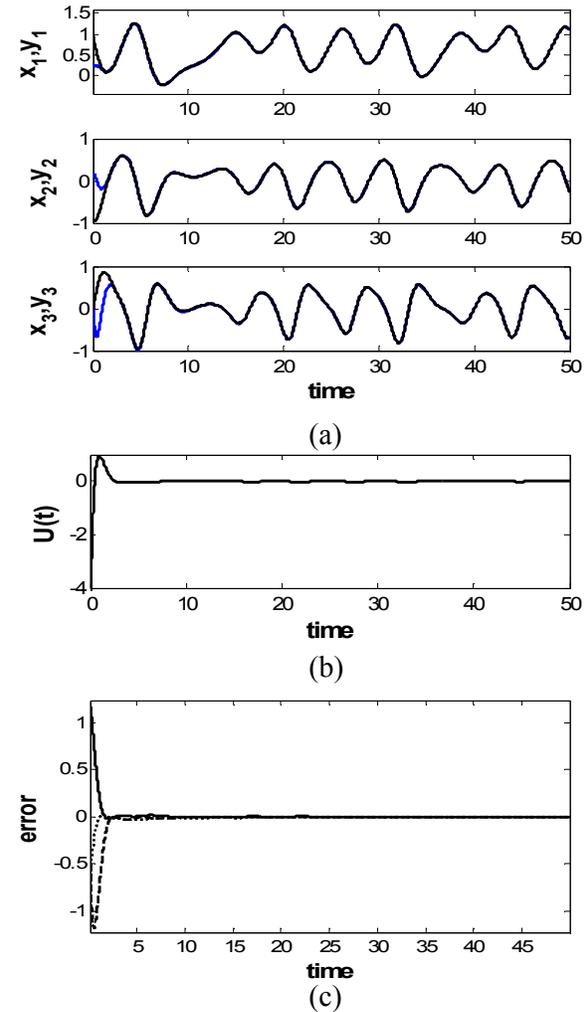

(a)

(b)

(c)

Figure 4: a) Simulation Results of synchronization of two identical fractional order Coullet systems with unknown parameter $b, c$ and $q = 0.95$ b) Control signal $u(t)$. c) The synchronization error.

$$u(t) = -\left(3\left(D^{1-q}\right)^3 + a + dy_1^2 + dx_1 e_1 + 2dx_1^2\right)e_1$$
$$-\left(5\left(D^{1-q}\right)^2 + \hat{b}\right)e_2 - \left(3\left(D^{1-q}\right) + \hat{c}\right)e_3 \quad (32)$$

$$\dot{\hat{b}} = -\left(2e_1 + 2D^{1-q}e_2 + \left(D^{1-q}\right)^2 e_3\right)\left(D^{1-q}\right)^3 x_2$$
$$\dot{\hat{c}} = -\left(2e_1 + 2D^{1-q}e_2 + \left(D^{1-q}\right)^2 e_3\right)\left(D^{1-q}\right)^3 x_3$$
(33)

## 5. CONCLUSION

In this paper, synchronization of two fractional order chaotic Coullet systems (as master and slave) with the same order is investigated. The advantage of the approach is that it only needs a single controller signal to perform the synchronization task. The synchronization will be achieved even in presence of unknown parameters by choosing a proper adaptation law. It is also shown that the adaptive backstepping procedure is feasible to construct the control law, using a recursive approach. Consequently, the extension of the control strategy to higher-dimensional hyperchaotic systems is promising. The simulation result verifies the significance of the proposed synchronization procedure in addition of the feasibility of control signals.